# Magnetless Circulators Based on Synthetic Angular-Momentum Bias: Recent Advances and Applications

Ahmed Kord, *Student Member, IEEE,* and Andrea Alù, *Fellow, IEEE*

*Abstract*— In this paper, we discuss recent progress in magnet-free non-reciprocal structures based on a synthetic form of angular momentum bias imparted via spatiotemporal modulation. We discuss how such components can support metrics of performance comparable with traditional magnetic-biased ferrite devices, while at the same time offering distinct advantages in terms of reduced size, weight, and cost due to the elimination of magnetic bias. We further provide an outlook on potential applications and future directions based on these components, ranging from wireless full-duplex communications to metasurfaces and topological insulators.

## I. INTRODUCTION

THE demand for multimedia-rich applications and services has been dramatically increasing over the last few decades. As a result, the sub-6 GHz electromagnetic spectrum, where most of deployed communication systems operate, has become saturated with unprecedented data traffic leaving no more bandwidth for further growth in the transmission rate. In order to overcome this problem, full-duplex radios have been proposed. In such systems, data streams are transmitted in the uplink and downlink simultaneously on the same carrier frequency [1]-[9], which doubles the spectral efficiency compared to half-duplex time- and frequency-division systems. The key challenge in full-duplexing, however, is that it requires considerably large self-interference cancellation (SIC) between the transmitter (Tx) and receiver (Rx) of every communication node in order to prevent the strong Tx signal from saturating the Rx frontend modules. For instance, WiFi signals are transmitted at an average power of +20 dBm and the noise floor is around –90 dBm, hence SIC is required to be as high as +110 dB. For many years, this level of SIC was impossible to attain, therefore full-duplex radios were presumed irrelevant in practice. Recently, several works challenged this long-held assumption and demonstrated that full-duplexing can, in fact, be achieved by using a combination of radio-frequency (RF) [3]-[5], analog [6]-[7], and digital techniques [8]-[9]. The basic idea in digital and analog approaches is to generate a weighted version of the Tx signal that mimics its echo into the receiver and subtract it from the Rx signal before and after the analog-to-digital converter. Since the strength contrast between the Tx and Rx signals can be billions in magnitude, the subtraction must be performed extremely accurately, otherwise it may end up adding more interference. But the presence of non-idealities such as noise, non-linearity, and dynamic variation of the antenna input impedance, make this process a non-trivial task. Therefore, analog and digital SIC approaches must be preceded by another layer of isolation at the RF frontend in order to relax their design requirements and mitigate their sensitivity to inevitable errors. Fortunately, circulators can play this role [10]-[64]. Not only they can achieve low insertion loss, thus overcoming the fundamental limit of electrical-balance duplexers, but they also allow the Tx and Rx nodes to share a single antenna thus making full-duplexing more attractive compared to multi-input multi-output (MIMO) radios.

For decades, circulators have been built almost exclusively through magnetic biasing of rare-earth ferrites [10]-[19]. These magnetic devices can achieve excellent performance in terms of isolation, power handling, bandwidth of operation, and insertion loss, yet they typically result in bulky, non-integrable, and expensive devices that may not be suited for various commercial systems. In order to overcome these problems, magnet-free implementations of circulators based on active solid-state devices or nonlinear materials with geometric asymmetries have been pursued [20]-[29]. Despite significant research efforts over many years, these alternative solutions continued to suffer from a fundamentally poor noise figure and limited power handling, hence the appeal of a fully passive material that can provide non-reciprocity kept favoring magnetic approaches. Recently, however, it was shown that parametric circuits can overcome these limitations and achieve excellent performance together with low-cost and small-size. For example, [31] introduced the concept of spatiotemporal modulation angular-momentum (STM-AM) biasing to synthesize cyclic-symmetric magnetless circulators. This was then followed by numerous works extending these concepts to

Manuscript received April 6, 2020. This work was supported by a Qualcomm Innovation Fellowship, an IEEE Microwave Theory and Techniques Society Graduate Fellowship, the Air Force Office of Scientific Research, the Defense Advanced Research Projects Agency, Silicon Audio, the Simons Foundation, and the National Science Foundation. (*Corresponding author: Andrea Alù.*)

A. Kord is with the Department of Electrical Engineering, Columbia University, New York, NY 10027, USA.

A. Alù is with the Department of Electrical and Computer Engineering, University of Texas at Austin, Austin, TX 78712 USA, and also with the Advanced Science Research Center, City University of New York, New York, NY 10031, USA. (e-mail: aalu@gc.cuny.edu)

Color versions of one or more of the figures in this paper are available online at http://ieeexplore.ieee.org.

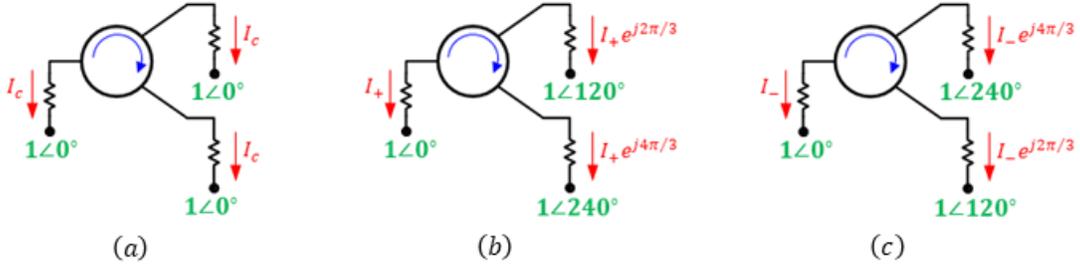

Fig. 1. Eigenmodes of a generic three-port circulator [35]. (a) In-phase mode. (b) Clockwise mode. (c) Counterclockwise mode.

realize various magnet-free components, including circulators, gyrators, and isolators [31]-[55]. In particular, [33]-[36] refined the STM-AM concept by finding the necessary conditions to ideally mimic magnetic biased cavities, thus enabling the development of many new circuits with unprecedented performance. In this article, we review recent advances in STM-AM devices and provide an outlook on their possible impact on technology and future research directions. We also discuss the unusual properties of two-dimensional metasurfaces consisting of arrays of such elements, which may be engineered to behave as magnetized ferrites or topological insulators [66]-[85].

## II. OPERATION PRINCIPLES OF MAGNETIC CIRCULATORS

The S-matrix of an ideal circulator can in general be written in the following form

$$\bar{\bar{S}} = \begin{bmatrix} 0 & 0 & 1 \\ 1 & 0 & 0 \\ 0 & 1 & 0 \end{bmatrix}, \quad (1)$$

Equation (1) shows that an input signal at, say, port 1 is exclusively routed to port 2 without any reflections or leakage to port 3. Similarly, an input signal at port 2 or 3 is routed to port 3 and 1, respectively. In other words, an ideal circulator is cyclic-symmetric and exhibits zero insertion loss, perfect matching, and infinite isolation. In reality, however, not only the S-matrix is limited by the inevitable dissipation in the constituent materials of the circulator, but it is also frequency dispersive and maintains non-reciprocity over a finite bandwidth (BW). Despite such practical imperfections, the harmonic response of a linear non-dispersive cyclic-symmetric circulator can still give us an invaluable insight into the physical principles of breaking reciprocity. For example, an excitation at one port of an ideal circulator, say port 1, can be decomposed using superposition into three sub-circuits, as shown in Fig. 1(a)-(c). In Fig. 1(a), the applied voltage sources are identical in both magnitude and phase while in Fig. 1(b) and Fig. 1(c), the sources have different phases, which increase by 120 deg either in the clockwise or counter-clockwise direction. Because of such distinctive phase pattern, we refer to such excitations as the in-phase (0), clockwise (+), and counter clockwise (−) modes, respectively. In general, each of these modal excitations generates a current component at the output ports and their linear combination gives the solution to the original excitation at port 1. If the circuit is symmetric and has only three terminals (no internal connections to ground), then a non-zero in-phase current violates Kirchhoff's current law, hence it cannot be excited. Furthermore, if a preferred sense of precession is provided to the counter-rotating currents, then they will be excited with an equal amplitude, say $I_g$, and opposite phase, say $\pm\alpha$. Consequently, the total current at the n-th port becomes

$$I_n = 2I_g \cos\left((n-1)\frac{2\pi}{3} + \alpha\right), \quad (2)$$

where $n$ is the port index. For $\alpha = 30$ deg, Equation (2) results in $I_3 = 0$ and $I_1 = -I_2 = \sqrt{3}I_g$. In other words, the impinging signal at port 1 is exclusively routed to port 2 and isolated from port 3. This implies that it is possible to implement an ideal circulator by imparting a preferred sense of precession to a pair of counter-roting modes in a symmetric three-port network.

This principle indeed governs the way in which magnetic circulators operate: consider a ferrite cavity connected symmetrically to three ports at 120 deg intervals as depicted in Fig. 2. If a strong magnetic bias is applied orthogonally to the cavity in order to align the spinning electron dipole moments along a particular direction, the spin imparts a form of angular-momentum bias at the microscopic level, which in turn translates into a preferred sense of precession for the rotating modes of the cavity. As a result, if these modes are excited by an impinging signal at one port, say port 1, they accumulate different phases as they appear the other two ports. For a particular strength of the bias, these modes can sum up in phase at one of the output ports, say port 2, and destructively interfere at the other, i.e., port 3. If power dissipation inside the cavity is negligible, then this means that the incident power is exclusively routed from port 1 to 2. Due to the geometrical three-fold symmetry of the circulator, excitation at ports 2 or 3, will similarly be routed to ports 3 and 1, respectively, which yields the S-matrix of (1).

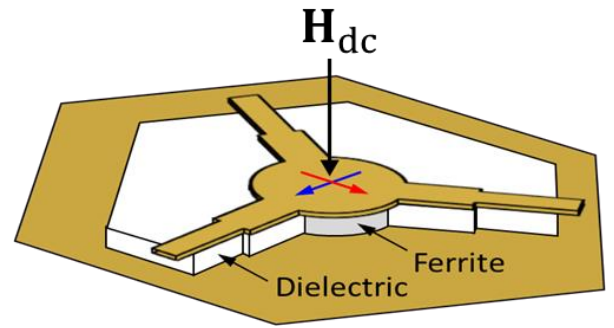

Fig. 2. Stripline Y-junction magnetic circulator [10].

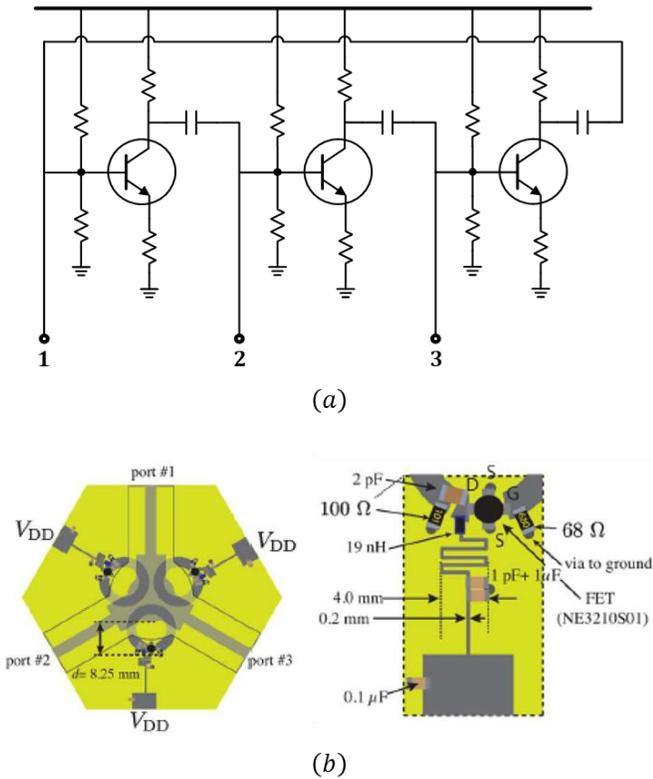

Fig. 3. (a) Active circulator consisting of three common-emitter amplifiers in a loop, each acting as an isolator [20]. (b) Active circulator consisting of three coupled transistor-loaded ring resonators forming an artificial metamaterial nonreciprocal core [22].

TABLE I
COMPARISON BETWEEN DIFFERENT CIRCULATOR DESIGN APPROACHES.

| Approach | Magnetic | Active | Parametric |
|---|---|---|---|
| Magnet | Required | Not required | Not required |
| Size | Large | Small | Medium |
| Cost | High | Low | Low |
| Insertion loss | Low | Low (or gain) | Medium |
| Isolation | Medium | Medium | High |
| Bandwidth | Wide | Average | Average |
| Power handling | High | Low | Medium |
| Noise figure | Low | High | Medium |
| Power dissipation | Low | High | Medium |

## III. LIMITATIONS OF ACTIVE CIRCULATORS

Despite the fact that magnetic-circulators can achieve excellent S-parameters, power handling and noise performance with negligible power consumption, they suffer from large size and high cost due to the fact that they require an external bulky magnet and rare-earth ferrite materials. These critical problems prohibited an ubiquitous use of such devices in wireless communication systems and played an important role in consolidating the false assumption that full-duplex radios are impossible to realize in practice. For this reason, an interest in investigating magnet-free approaches to design circulators has emerged since the advent of solid-state transistors. When a BJT transistor, for instance, is biased with a dc current, it operates as a trans-conductance amplifier which converts ac voltages applied at its base terminal into large currents flowing from its emitter to collector terminals. A reverse voltage applied at either of the output terminals, however, is isolated from the base. In other words, a current-biased transistor operates as an isolator with forward gain. By combining three of such active isolators in a loop as shown in Fig. 3(a), an active magnet-free circulator can be constructed as explained in [20]. Variants of this circuits at different frequency ranges were also investigated over the years. For example, a robust low-frequency implementation using operational amplifiers (op-amps) was also presented in [21]. More recently, a synthetic active metamaterial consisting of an array of transistor-loaded microstrip ring resonators was also proposed in [22] as an artificial medium to replace magnetic-biased ferrites. In such materials, the transistors simply kill one of the rotating modes supported by the ring, thus allowing to realize active circulators as depicted in Fig. 3(b).

It is also worth mentioning that nonlinear materials have also been explored as an alternative medium to realize magnet-free circulators, especially at optical frequencies where transistors do not exist [27]-[29]. However, the resulting devices were only able to work when only one port is excited at a time. When two or more ports are concurrently excited, the S-parameters suffer from hysteresis and strongly depend on the relative phase between the input signals. Therefore, this approach is less useful than its active transistor-based counterpart which can support simultaneous excitation from different ports and at the same time offers distinct advantages in terms of weight, size, cost, and scalability compared to magnetic-biased ferrite devices. Unfortunately, however, active circulators still suffer from low power handling and poor noise figure. A comprehensive analysis of these fundamental limitations was presented in [26] leading to the conclusion that active circulators are not suitable for use as frontend modules in the vast majority of microwave applications where Tx power is either high or Rx sensitivity is critical.

## IV. PARAMETRIC STM-AM CIRCULATORS

In recent years, time-varying circuits were presented as a promising solution to realize circulators that could handle much higher power, exhibit better linearity, and achieve lower noise figure than active implementations, while at the same time eliminating the necessity for bulky magnets and expensive materials needed in ferrite devices. Table I summarizes these advantages in comparison to traditional magnetic and active technologies presented in previous sections. Among all recent works on parametric circulators, STM-AM circuits, which is the main focus of this article, have shown significant promise in satisfying the challenging requirements of practical systems at different portions of the electromagnetic spectrum. In particular, our group presented in [56] a circulator for airborne sound by synthesizing an effective angular momentum bias, in the form of mechanical rotation of air, inside a circularly symmetric resonant acoustic cavity. Such bias imparted a macroscopic sense of precession to the degenerate modes of opposite handedness supported by the cavity, which translates into non-reciprocity and isolation as explained in Sec. II. Using similar principles, one could imagine spinning a circuit board sufficiently fast to break reciprocity for the electromagnetic signals traveling across a resonant device, but arguably this mechanical spin would be even less practical than magnetic approaches to non-reciprocity. However, spatiotemporal modulation can impart a form of synthetic angular momentum bias, enabling the implementation of compact STM-AM circulators as described in the following.

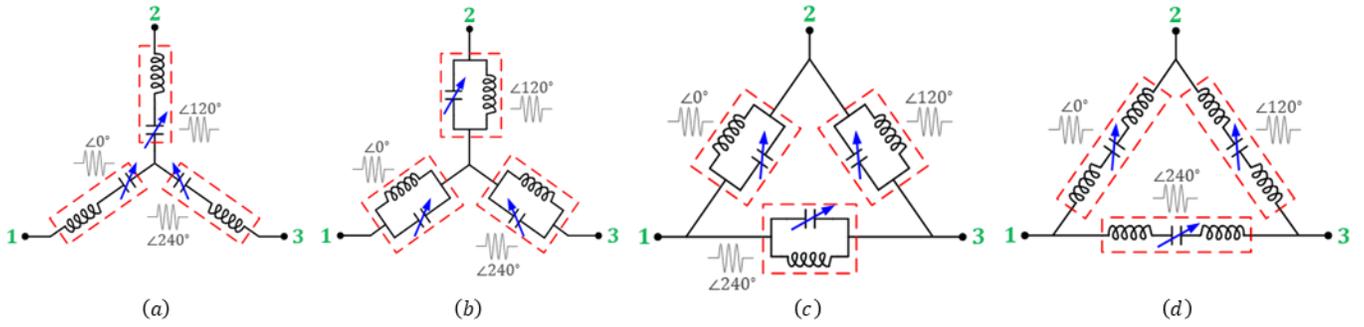

Fig. 4. Single-ended topologies of STM-AM circulators [35]. (a) Bandpass/wye. (b) Bandstop/wye. (c) Bandstop/delta. (d) Bandpass/delta.

*A. Single-Ended Topologies*

Consider a symmetric resonant three-port network consisting of three series or parallel $LC$ tanks connected in a loop or to a central node, resulting in four possible combinations as shown in Fig. 4. Let us further assume that the instantaneous oscillation frequencies of the tanks are spatiotemporally modulated as follows

$$f_n = f_0 + \Delta f \cos\left(2\pi f_m t + (n-1)\frac{2\pi}{3}\right), \qquad (3)$$

where $f_0 = \frac{1}{2\pi\sqrt{L_0 C_0}}$ is the natural oscillation frequency of the tanks, $\Delta f$ is the modulation depth, and $f_m = \omega_m/2\pi$ is the modulation frequency. In practice, (3) is implemented using varactors, switched capacitors or Josephson junctions, depending on the target application. In the case of varactors, for instance, $\Delta f$ can be written as $\Delta f = kV_m$ where k is the slope of the C-V curve. The modulation scheme of (3) synthesizes an effective angular momentum bias in the clockwise direction since the phases of the modulation signals increase by 120 deg in that direction thus lifting the degeneracy of the rotating modes and allowing them to oscillate at different frequencies, say $\omega_\pm$. Between these two frequencies, the skirts of the resonances do allow transmission, whether the original resonance was bandpass or bandstop, and they exhibit opposite phases. Therefore, if an input signal is incident at one port in the middle of these two frequencies, i.e., at $\omega_0$, it will excite the two resonances with equal magnitude and opposite phases $\pm\alpha$, which is consistent with the generic description of the previous section based on providing a preferred sense of precession. It is also worth mentioning that the amount of splitting $\omega_+ - \omega_-$ depends on the modulation parameters $f_m$ and $V_m$, thus allowing to control the value of $\alpha$. As explained earlier, for $\alpha = 30$ deg, ideal circulation can be achieved at $\omega_0$. In light of this discussion, this condition will correspond to a certain combination of $f_m$ and $V_m$.

As an example, the bandstop/delta topology of Fig. 4(c) was built and tested in [33]. The measured insertion loss (IL), return loss (RL), and isolation (IX) at the center frequency of 1 GHz were 3.3 dB, 10.8 dB, 55 dB, respectively, and the measured BW was 2.4% (24 MHz). This circuit can also be reconfigured for operation at different channels over a frequency range of 60 MHz by simply changing the DC bias of the varactors. The measured 1- and 20-dB compression points of IL and IX, respectively, were both +29 dBm. Also, the measured IIP3 was +33.8, which are reasonable nonlinearity metrics meeting the specifications for a range of applications. It is important to stress that these metrics stem from the properties of the employed circuit components, not from a fundamental limit in the approach to non-reciprocity. It can also be proven that STM-AM circulators are inherently passive linear circuits [42], hence we expect that the overall noise figure (NF) is approximately equal to the IL but slightly larger due to noise folding from the intermodulation products (IMPs) and phase noise in the modulation signals. Indeed, the measured NF was 4.5 dB at 1 GHz and less than 4.7 dB over the BW. These results are summarized in Table II and the interested reader is referred to [33] for further technical details.

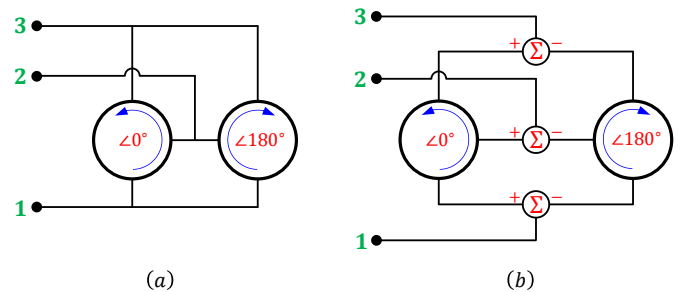

Fig. 5. Differential configurations of STM-AM circulators [34]. (a) Current-mode. (b) Voltage-mode.

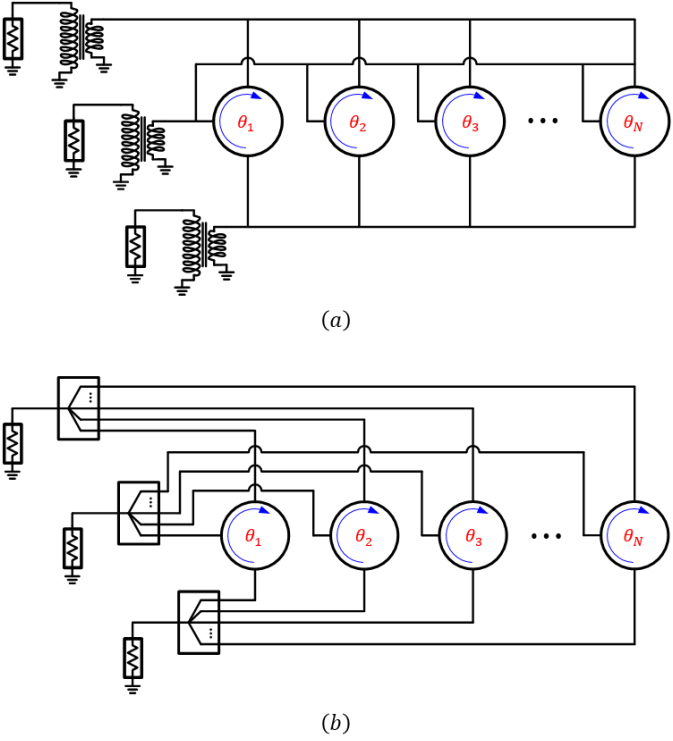

(a)

(b)

Fig. 6. *N*-way architectures of STM-AM circulators [35]. (a) Parallel interconnection based on current combination. (b) Series interconnection based on voltage summation.

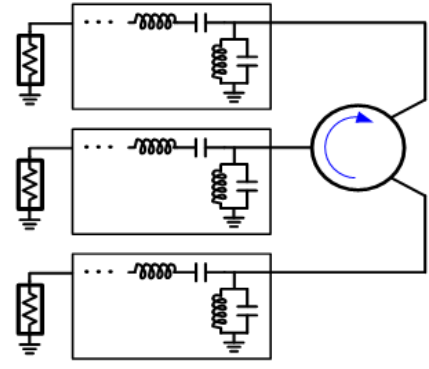

Fig. 7. Bandwidth extension of STM-AM circulators based on terminating the ports of a narrowband differential or N-way circuit with three properly designed matching filters.

### B. Differential Configurations

As illustrated in previous section, single-ended STM-AM circulators can achieve good performance in terms of several metrics, particularly in terms of power handling and noise figure. Nevertheless, these circuits suffer from large intermodulation products (IMPs) in close proximity to the desired BW, resulting from parametric mixing induced by the temporal modulation and from the inevitable non-linearities associated with its circuit components. These IMPs draw power from the input signal, which limits IL to about 2 dB in practice. They are also considerably large, i.e., only 10 dB below the fundamental harmonic, and they equally appear at all ports for excitation at a single port. Therefore, the IMPs resulting at the Rx port from the Tx excitation could saturate the Rx front-end regardless of how much isolation is achieved at the fundamental frequency. Similarly, the IMPs resulting from the Rx signal at the Tx port can cause instability issues to the power amplifier (PA) because of load-pull effects. More importantly, the IMPs at the antenna (ANT) port, either due to the Tx or the Rx signal, pose a serious interference problem to adjacent channels and would prohibit compliance with the spectral mask regulations of most commercial systems. Therefore, it is crucial to reject these products. This problem was overcome in [34] by combining two single-ended circuits similar to those presented in the previous section in a differential voltage- or current-mode architecture, as shown in Fig. 5 and maintaining a constant 180 deg phase difference between their modulation signals, i.e.,

$$f_{n,\pm} = f_0 \pm \Delta f \cos\left(2\pi f_m t + (n-1)\frac{2\pi}{3}\right) \quad (4)$$

where $f_{n,+}$ and $f_{n,-}$ are the modulation signals of the two constituent single-ended circuits. As an example, in [34], the voltage-mode configuration of Fig. 5(b) was built and tested using two of the single-ended bandstop/delta circuits depicted in Fig. 4(c). The measured IL, RL, and IX after de-embeding the baluns were 0.78 dB, 23 dB, and 24 dB, respectively, and the fractional BW was 2.3% (23 MHz). Table II also summarizes these results along with other metrics and the interested reader is referred to [34] for further details.

### C. N-Way Architectures

While the differential STM-AM circulators presented in previous section dramatically improved the performance over many metrics compared to the single-ended implementations, their spurious emission can remain strong when non-idealities in the circuit are considered. In order to address this problem, we can extend the differential configuration to an *N*-way architecture, consisting of *N* unit cells connected in parallel or in series as shown in Fig. 6 [35]. The modulation scheme in this case becomes

$$f_{ni} = f_0 + \Delta f \cos\left(2\pi f_m t + (n-1)\frac{2\pi}{3} + (i-1)\frac{2\pi}{N}\right) \quad (5)$$

where $i = 1{:}N$ is the unit index and $n = 1{:}3$ is the tank index in the *i*-th unit. Equation (5) reduces to (4) for $N = 2$, and this generalization is more effective at eliminating the IMPs that are not multiples of $Nf_m$ in the presence of nonidealities, such as non-linearities, phase synchronization errors and random mismatches between the constituent elements. As a by-product, *N*-way circulators provide better power handling, enhanced by a factor of *N* since the input power is split amongst *N* units. Nonetheless, the benefits of the N-way architectures come at the expense of an overall increased size and power consumption. As before, Table II summarizes the results and the interested reader is referred to [35] for further technical details.

### D. Bandwidth Extension

The instantaneous BW of all circuits presented thus far is limited by the modulation frequency $f_m$, the loaded quality factor $Q_l$ of the resonant junction, and the order of the constituent resonators. For example, first-order resonators, i.e., series or parallel *LC* tanks, with 10~20% modulation frequency and 50 Ohm termination led to a BW of about 3~4% at best. One solution to improve this further is to increase $f_m$, but this approach increases power consumption, prohibits the use of thick-oxide low-speed technologies that could handle high power, and complicates scaling the center frequency to the mm-wave bands. Alternatively, one may think of replacing the LC tanks with high-order LC filters but this would increase the circuit complexity and size dramatically. In [36], an optimal solution based on terminating a narrowband circulator with three identical bandpass filters, as depicted in Fig. 7, was proposed. The added filters are essentially matching networks that shape the frequency of the combined network and maintain a certain level of IX over a wider BW. A theoretical bound on the maximum possible bandwidth for a given modulation frequency was also derived in [36]. As one may expect, in order to approach such a bound, the order of the matching filters must be very large. In practice, however, the maximum order is limited by the additional losses exhibited by the filters themselves which increase the circulator's overall IL. In order to experimentally validate this technique, [36] relied on a 2nd-order filter and a current-mode bandpass/wye STM-AM circuit resulting in a three-fold increase in the BW from 4% (40 MHz) to 14% (140 MHz). The measured results of all other metrics are again summarized in Table II and the interested reader is referred to [36] for further details.

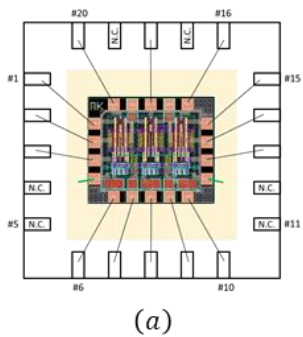 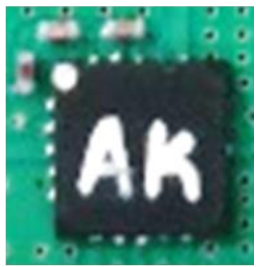

Fig. 8. Photographs of the first CMOS STM-AM circulator [37]. (a) Chip. (b) Board.

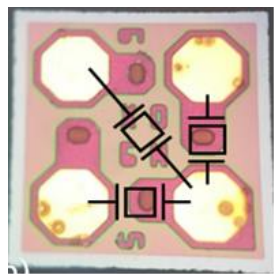 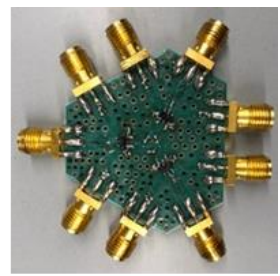

Fig. 9. Photographs of the first MEMS STM-AM circulator [41]. (a) Chip. (b) board.

### E. Integrated Implementations

The results presented in the previous sections were all based on printed circuit board (PCB) prototypes which, despite outperforming previous works in many metrics, were still limited in size. To overcome this problem and to achieve significant miniaturization, an IC implementation is highly desirable. Towards this goal, [37] modified the bandpass/wye

TABLE II
SUMMARY OF THE MEASURED RESULTS IN COMPARISON TO PREVIOUS WORKS.

| Reference | [33] | [34] | [35] ¥ | [36] | [37] | [41] |
|---|---|---|---|---|---|---|
| Architecture | Single-ended Bandstop/Delta | Differential Bandstop/Delta | Parallel N-way Bandstop/Delta | Broadband Current-mode Bandpass/Wye | Current-mode Bandpass/Wye | Current-mode Bandpass/Wye |
| Technology | PCB | PCB | N/A | PCB | CMOS 180-nm | MEMS/PCB |
| Cent. frequency (GHz) | 1 | 1 | 1 | 1 | 0.91 | 2500 |
| Mod. frequency (%) | 19 | 10 | 10 | 11 | 11.6 | 1.6 |
| 20 dB IX BW (%) | 2.4 | 2.3 | 3 | 14 | 2.4 | 0.72 |
| Max IX (dB) | 55 | 24 | 25 | 31 | 65 | 30 |
| Min IL (dB) | 3.3 | 0.8 † | 2.2 | 4.2 | 4.8 | 4 |
| Max RL (dB) | 11 | 23 | 24 | 16.7 | 11 | 15 |
| P1dB (dBm) | +29 | +29 | +39 | +23 | N/A | +28 |
| IX20dB (dBm) | +29 | +28 | +41 | N/A | N/A | N/A |
| IIP3 (dBm) | +33 | +32 | +43 | N/A | +6 | +40 |
| Min NF (dB) | 4.5 | 2.5 | 2.2 | N/A | 5.2 | N/A |
| Max. IMP (dBc) | −11 @ $f_0 \pm f_m$ | −30 @ $f_0 \pm f_m$ | −62 @ $f_0 \pm 2f_m$ | −22 @ $f_0 \pm f_m$ | −20 @ $f_0 \pm f_m$ | −30 @ $f_0 \pm f_m$ |
| Power diss. (mWatt) | N/A | N/A | N/A | N/A | 64 | 0.15 |
| Size (mm²) | 143 | 286 | N/A | 480 | 36 | 225 * |

N/A: Not available.    ¥ Simulation results.    † Baluns are de-embedded.    * MEMS die area = 0.25 mm².

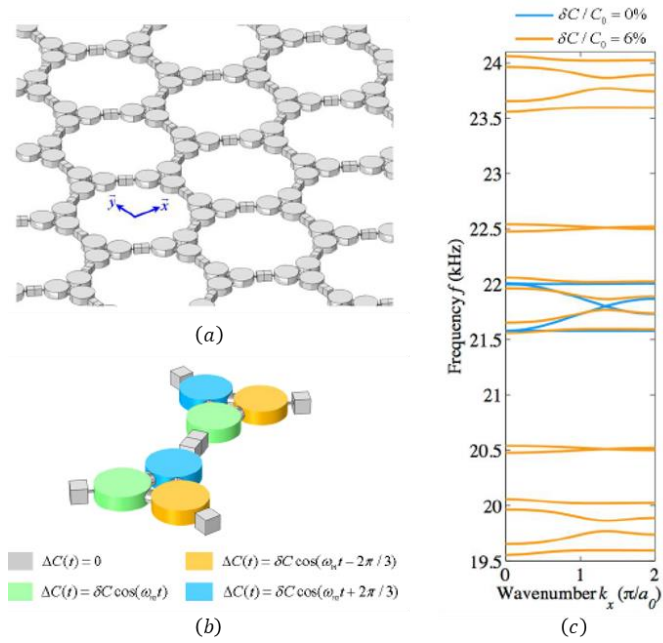

Fig. 10. (a) Hexagonal lattice consisting of acoustic STM-AM circulators forming a Floquet topological insulator for sound waves [81]. (b) Unit cell consisting of two acoustic STM-AM circulators. (c) Comparison between the band structure of an infinite lattice with and without modulation showing a band gap opening in the vicinity of 22 MHz when modulation is turned on.

topology of Fig. 4(a) to be compatible with CMOS technologies and experimentally validated the resulting circuit in a current-mode architecture using a 180nm process. In particular, the modulated capacitors were implemented using switched capacitors which result in a larger modulation index than varactors, alleviate the tradeoff between the index and the quality factor, and permit the use of digital clocks which can be easily generated on chip. Fig. 8 shows photographs of the fabricated chip and the testing board presented in [37]. It is also worth mentioning that this was the first CMOS implementation of STM-AM circulators ever-presented. The measured RL, IL, and IX at the center frequency of 910 MHz were about 12 dB, 4.9 dB, and 60 dB, respectively. Also, the measured IIP3 was +6 dBm which was limited by the ESD protection circuitry and the NF was 5.2 dB. Similarly, these results are all summarized in Table II and the interested reader is referred to [37] for further details.

While a CMOS implementation of STM-AM circulators does indeed reduce the size significantly compared to a PCB design, extreme miniaturization is still limited by the fact that on-chip inductors consume large silicon area and exhibit low quality factors. This problem can be overcome by using micro-machined integrated devices such as MEMS resonators. Such devices can exhibit a series resonance with a very high Q which mimics the response of a series LC tank while being much smaller in size. However, MEMS resonators also suffer from a parasitic parallel resonance in close proximity to its series resonance. For devices operating in the low GHz range, the parasitic resonance can be shifted to higher frequencies with a small inductor of 0.5~1 nH which can be realized using a bondwire thus maintaining the benefit of reducing the circulator's area. As an example, [41] presented the first MEMS implementation of STM-AM circulators as depicted in Fig. 9. The measured RL, IL, and IX at the center frequency of 2.5 GHz were about 15 dB, 4 dB, and 30 dB, respectively. A distinct advantage of MEMS implementations is also manifested by the fact that they allow a significant reduction of the modulation frequency. In this design, the modulation frequency was as small as 40 MHz (1.6%) which resulted in a very low power consumption of only 150 µW. However, this advantage comes at the expense of a narrowband bandwidth which was about of 18 MHz (0.72%). These results are summarized in Table II and the interested reader is referred to [41] for further details.

## V. METAMATERIAL APPLICATIONS

Applications of the STM-AM concepts presented thus far have been heavily investigated over the last few years. While full-duplex communication is an obvious example, as explained earlier in this article, other applications have also been explored. Specifically, in the field of metamaterials STM-AM biasing has been recently exploited to synthesize Faraday rotation for free-space propagating waves hitting a metasurface formed by arrays of such elements, and topological phenomena for surface wave propagation along such arrays. A magneto-optical material biased by a DC magnetic field supports circular bi-refringence and Farday rotation, which may be used as a way to realize non-reciprocal wave propagation for free-space radiation. Similarly, an array of STM-AM elements like those described in the previous sections can impart circular birefringence over a thin surface, causing Faraday rotation to a linearly polarized impinging plane wave that may be directly translated in the implementation of free-space isolators [30]. The non-reciprocal polarization rotation imparted by such metasurfaces can reach 1,000s of degrees per wavelength, a very impressive metric considering that these surfaces can be implemented in standard CMOS technology without need of an applied magnetic bias. Beyond electro-optical modulation of these arrays, similar phenomena can be achieved using all-optical modulation via nonlinearities. By parametrically pumping a nonlinear etalon with two circularly polarized waves at detuned frequencies, it is possible to impart an effective angular momentum bias that replaces the modulation schemes discussed in the previous sections [65].

Opportunities arise also to impart non-reciprocal responses to surface waves propagating along an array of STM-AM biased devices. Arguably a striking example of this type is the opportunity to mimic the response of a topological insulator (TI) for electromagnetic waves. A TI is a material that acts as an insulator in the bulk and as a conductor on its edge or boundary, independent of the way in which the boundary is deformed or changed [66]-[69]. What is unique about such materials is that its edge conduction states are symmetry-protected by the topology of their conduction bands delimiting the bandgap, hence they allow unusually robust conduction properties at the foundations of the quantum Hall effect, among many other unusual phenomena. Translating these concepts from electronics to electromagnetics, a photonic TI enables unidirectional reflection-free propagation on the edges of the (meta)-material witnin a photonic bandgap, immune against a wide range of structural imperfections. The existence of such modes can be directly predicted from the topology of the bands of an infinite lattice, by calculating the so-called topological invariant, a quantity that describes the nature of the bandgap boundaries in momentum-frequency space. If the topological invariant of the bands delimiting the bandgap are different, then the bandgap necessarily supports a boundary mode, which is inherently topologically protected from disorder and back reflections.

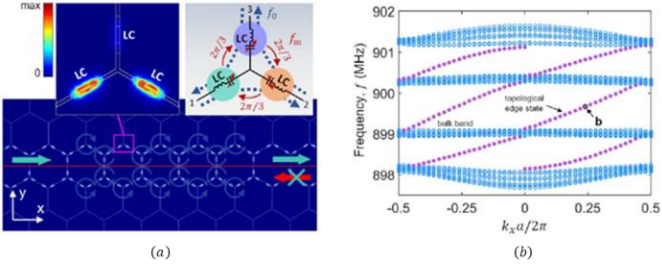

Fig. 11. (a) Hexagonal lattice consisting of electronic STM-AM circulators based on the bandpass wye topology of Fig. 2(a) [85]. An artificial boundary is created at the middle of two domains inside the lattice, each with an opposite direction of STM-AM bias. Schematic and energy density of a single meta-atom near the artificial interface are shown in the insets. (b) Band diagram with four conduction bands (blue) and a topologically protected edge state (purple) at the artificial interface.

While TIs were originally envisioned and discovered in condensed-matter physics, their extension to classical photonic [69]-[78], acoustics [79]-[83], and circuits [83]-[85] has been driving a lot of excitement. In such a case, symmetry breaking at a degeneracy point in momentum space is typically sufficient to induce topological order and enable edge states. Breaking time-reversal symmetry is a particularly effective way to realize electromagnetic TIs, since it supports robust, broadband non-reciprocal propagation and guarantees the absence of reflected modes regardless of the nature of the possible disorder in the system.

Here, we consider an array of STM-AM elements that realize an electromagnetic TI with non-reciprocal edge propagation. Since the edge states in this case emerge from periodic temporal variations, the resulting system is referred to as a Floquet TI. An example is shown in Fig. 10(a) [81]: the unit cell of the corresponding hexagonal lattice is depicted in Fig. 10(b) and it consists of two waveguide STM-AM circulators in which the time modulation synthesizes a rotation, similar to what described in the previous sections, to impart an effective angular momentum bias. The STM-AM waveguide circulator is consistent with the bandpass/delta topology of Fig. 4(d), and the modulation can be achieved by modulating the volume of each cavity using piezoelectric actuators, or in other similar schemes as described in the previous sections. Fig. 10(c) shows a comparison between the band diagrams of such a photonic crystal with and without modulation. Without modulation (blue curves), four propagating bands are found around 22 kHz, two of which appear to be slow in nature corresponding to *deaf* modes, while the other two correspond to fast Dirac bands with degeneracy at both G and K points. On the other hand, when the modulation is turned on (orange curves), the band diagram folds with periodicity equal to the modulation frequency, as would be expected from the Floquet-Bloch theorem. More importantly, since the modulation breaks the temporal symmetry, the degeneracy at both G and K points is lifted and a bandgap opens. Ref. [81] theoretically predicted the existence of a unidirectional edge state in such a gap by calculating the topological invariant of the first Chern class for the four conduction bands. Such a prediction was then validated through full-wave simulations of a truncated lattice confirming the theoretical analysis.

In [85], an electronic STM-AM TI was also presented. Similar to the TI of Fig. 10, the lattice was also hexagonal but the meta-atom was based on a bandpass/wye STM-AM circulator, as in Fig. 4(a). Without modulation, the lattice exhibits a continuous energy spectrum with six Dirac points. When the STM-AM biasing is applied to each meta-atom, the degeneracy of the Dirac points is lifted and a bandgap opens in the bulk. Furthermore, in order to test the presence of an edge mode, the lattice can be divided it into two regions, one with a clockwise STM-AM bias and the other with a counter clockwise bias. Such a division results in a domain wall between the two regions that can support an edge state and is easier to calculate numerically. Fig. 11(a) shows the spatial distribution of the energy density of such a mode when excited at the middle of the bandgap. As depicted in the left inset, the energy is localized around the artificial boundary and decays sharply in the bulk. Fig. 11(b) shows the band diagram for such a scenario, which indeed mainfests an edge mode that can carry energy only in the +x direction. An experimental demonstration for elastic waves has also been recently reported using arrays of piezo-electric elements [82], also showing dynamic reconfigurability and opportunities for multiplexing.

STM-AM Floquet TIs do not require fast modulation speeds nor phase uniformity between different unit cells, which eases their implementation and makes them relevant for pratical applications, for instance for multiplexing. The channels can be dynamically reconfigured thanks to the possibility of creating artificial domain walls by simply controlling the direction of the STM-AM bias, which is of significant importance in today's programmable systems. This, in turn, opens the door to new venues and possibilities in future research directions.

## VI. Conclusions

In this paper, we reviewed the concept and recent progress on magnet-free STM-AM devices. We outlined the underlying physical principles and presented its basic topologies and advanced architectures, with a focus on the single device element to realize circulators. For completeness, the results of all fabricated prototypes are summarized in Table II. We also provided an outlook on the potential impact STM-AM circulators may have on future technologies such as wireless full-duplex communications and in the field of metamaterials such as Floquet topological insulators. The antennas and propagation community can play a pivotal role in pushing forward the application research on these systems, in which many outstanding challenges remain, both from the modeling and theoretical perspective, for instance in the efficient calculation of time-modulated systems, and from the practical and efficient implementation of these devices for various applications.

REFERENCES

[1] D. Korpi, J. Tamminen, M. Turunen, T. Huusari, Y. S. Choi, L. Anttila, S. Talwar, and M. Valkama, "Full-duplex mobile device: pushing the limits," IEEE Commun. Mag., vol. 54, no. 9, pp. 80–87, September 2016.

[2] A. Sabharwal, P. Schniter, D. Guo, D. W. Bliss, S. Rangarajan, and R. Wichman, "In-Band Full-Duplex Wireless: Challenges and Opportunities," IEEE J. Sel. Areas Commun., vol. 32, no. 9, pp. 1637–1652, Sept 2014.

[3] J. I. Choi et al., "Achieving single channel, full duplex wireless communication," in Proc. 16th Annu. Int. Conf. Mobile Comput. Netw. ACM, Chicago, IL, USA, Sept. 2010, pp. 1-12.

[4] M. Duarte, C. Dick, and A. Sabharwal, "Experiment-driven characterization of full-duplex wireless systems," IEEE Trans. Wireless Commun., vol. 11, no. 12, pp. 4296-4307, Dec. 2012.

[5] A. Kord, D. L. Sounas, and A. Alù, "Achieving Full-Duplex Communication: Magnetless Parametric Circulators for Full-Duplex Communication Systems," IEEE Microw. Mag., vol. 19, no. 1, pp. 84-90, Jan. 2018.

[6] B. Debaillie et al., "Analog/RF Solutions Enabling Compact Full-Duplex Radios," IEEE J. Sel. Areas Commun., vol. 32, no. 9, pp. 1662–1673, Sept 2014.

[7] J. Zhou et al., "Integrated full duplex radios," IEEE Commun. Mag., vol. 55, no. 4, pp. 142–151, Apr. 2017.

[8] M. Jain et al., "Practical, real-time, full duplex wireless," in Proc. 17th Annu. Int. conf. Mobile Comput. Netw. ACM, Las Vegas, NV, USA, Sept. 2011, pp. 301-312.

[9] D. Bharadia et al., "Full duplex radios," ACM SIGCOMM Comput. Commun. Rev., vol. 43, no. 4, pp. 375-386, Sept. 2013.

[10] D. M. Pozar, Microwave engineering, John Wiley & Sons, 2009.

[11] A. Kord, D. L. Sounas and A. Alù, "Microwave Nonreciprocity," in *Proceedings of the IEEE*, 2020.

[12] N. Reiskarimian *et al.*, "Nonreciprocal electronic devices: A hypothesis turned into reality," IEEE Microwave Magazine, Vol. 20, no. 4, pp. 94-111, 2019.

[13] C. Caloz and A. Sihvola, "Electromagnetic Chirality, Part 1: The Microscopic Perspective [Electromagnetic Perspectives]," IEEE Antennas and Propagation Magazine, Vol. 62, no. 1, pp. 58-71, 2020.

[14] S. Taravati and A. A. Kishk, "Space-time modulation: Principles and applications," IEEE Microwave Magazine, Vol. 21, no. 4, pp. 30-56, 2020.

[15] J. A. Weiss, N. G. Watson, and G. F. Dionne, "New uniaxial-ferrite millimeter-wave junction circulators," in Proc. Int. Symp. IEEE MTT-S Microw., Long Beach, CA, USA, 1989, vol. 1, pp. 145–148.

[16] A. Saib, M. Darques, L. Piraux, D. Vanhoenacker-Janvier, and I. Huynen, "Unbiased microwave circulator based on ferromagnetic nanowires arrays of tunable magnetization state," J. Phys. D, Appl. Phys., vol. 38, no. 16, pp. 2759–2763, Aug. 2005.

[17] L. P. Carignan, A. Yelon, D. Menard, and C. Caloz, "Ferromagnetic nanowire metamaterials: Theory and applications," IEEE Trans. Microw. Theory Techn., vol. 59, no. 10, pp. 2568–2586, Oct. 2011.

[18] S. A. Oliver et al., "Integrated self-biased hexaferrite microstrip circulators for millimeter-wavelength applications," IEEE Trans. Microw. Theory Techn., vol. 49, no. 2, pp. 385-387, 2001.

[19] J. Wang et al., "Self-biased Y-junction circulator at Ku band," IEEE Microw. Wireless Compon. Lett., vol. 21, no. 6, pp. 292-294, 2011.

[20] S. Tanaka, N. Shimomura, and K. Ohtake, "Active circulators: The realization of circulators using transistors," Proc. IEEE, vol. 53, no. 3, pp. 260-267, 1965.

[21] A. W. Keen, J. L. Glover and R. J. Harris, "Realisation of the circulator concept using differential-input operational amplifiers," Electron. Lett., vol. 4, no. 18, pp. 389-391, Sept. 1968.

[22] T. Kodera et al., "Magnetless nonreciprocal metamaterial (MNM) technology: application to microwave components," IEEE Trans. Microw. Theory Techn., vol. 61, no. 3, pp. 1030–1042, Mar. 2013.

[23] T. Kodera, D. L. Sounas, and C. Caloz, "Artificial Faraday rotation using a ring metamaterial structure without static magnetic field," Appl. Phys. Lett., vol. 99, Jul. 2011, Art. no. 03114.

[24] S. Wang, C. H. Lee, and Y. B. Wu, "Fully Integrated 10-GHz Active Circulator and Quasi-Circulator Using Bridged-T Networks in Standard CMOS," IEEE Trans. Very Large Scale Integr. (VLSI) Sys., vol. 24, no. 10, pp. 3184–3192, Oct. 2016.

[25] C.-H. Chang, Y.-T. Lo, and J.-F. Kiang, "A 30 GHz active quasicirculator with current-reuse technique in 0.18 μm CMOS technology," IEEE Microw. Wireless Compon. Lett., vol. 20, no. 12, pp. 693–695, Dec. 2010.

[26] G. Carchon and B. Nanwelaers, "Power and noise limitations of active circulators," IEEE Trans. Microw. Theory Techn., vol. 48, no. 2, pp. 316–319, Feb. 2000.

[27] P. Aleahmad, M. Khajavikhan, D. Christodoulides, and P. LiKamWa, "Integrated multi-port circulators for unidirectional optical information transport," Sci. Rep., vol. 7, p. 2129, 2017.

[28] K. Y. Yang, J. Skarda, M. Cotrufo, A. Dutt, G. H. Ahn, D. Vercruysse, S. Fan, A. Alù, and J. Vuckovic, "Inverse-designed photonic circuits for fully passive, bias-free Kerr-based nonreciprocal transmission and routing," Nature Photonics, in press, 2020.

[29] G. D'Aguanno, D. L Sounas, H. M. Sayed, and A. Alù, "Nonlinearity-based circulator," Applied Physics Letters, vol. 114, No. 18, 181102 (4 pages), May 7, 2019.

[30] D. L. Sounas, C. Caloz, and A. Alù, "Giant non-reciprocity at the subwavelength scale using angular momentum-biased metamaterials," Nature Commun., vol. 4, no. , pp. 2407, 2013.

[31] N. A. Estep et al., "Magnetic-free non-reciprocity and isolation based on parametrically modulated coupled-resonator loops," Nat. Phys., vol. 10, no. 12, pp. 923-927, Nov. 2014.

[32] N. A. Estep, D. L. Sounas, and Andrea Alù, "Magnetless Microwave Circulators Based on Spatiotemporally Modulated Rings of Coupled Resonators," IEEE Trans. Microw. Theory Techn., vol. 64, no. 2, pp. 502-518, Feb. 2016.

[33] A. Kord, D. L. Sounas, and A. Alù, "Magnetless Circulators Based on Spatio-Temporal Modulation of Bandstop Filters in a Delta Topology," IEEE Trans. Microw. Theory Techn., vol. 66, no. 2, pp. 911-926, Feb. 2018.

[34] A. Kord, D. L. Sounas, and A. Alù, "Pseudo-Linear Time-Invariant Magnetless Circulators Based on Differential Spatio-Temporal Modulation of Resonant Junctions," IEEE Trans. Microw. Theory Techn., vol. 66, no. 6, pp. 2731-2745, Jun. 2018.

[35] A. Kord, H. Krishnaswamy, and A. Alù, "Magnetless Circulators with Harmonic Rejection Based on N-Way Cyclic-Symmetric Time-Varying Networks," Phys. Rev. Appl., vol. 12, no. 2, pp. 024046(1-14), Aug. 2019.

[36] A. Kord, D. L. Sounas, Z. Xiao, and A. Alù, "Broadband Cyclic-Symmetric Magnet-less Circulators and Theoretical Bounds on their Bandwidth," IEEE Trans. Microw. Theory Techn., vol. 66, no. 12, Dec. 2018, pp. 5472-5481.

[37] A. Kord, M. Tymchenko, D. L. Sounas, H. Krishnaswamy, and A. Alù, "CMOS Integrated Magnetless Circulators Based on Spatiotemporal Modulation Angular-Momentum Biasing," IEEE Trans. Microw. Theory Techn., vol. 57, no. 7, pp. 2649-2662, July 2019.

[38] C. Yang and P. Gui, "85–110-GHz CMOS magnetic-free nonreciprocal components for full-duplex transceivers," IEEE J. Solid-State Circuits, vol. 54, no. 2, pp. 368-379, 2018.

[39] C. Cassella, G. Michetti, M. Pirro, Y. Yu, A. Kord, D. Sounas, A. Alù, and M. Rinaldi, "Radio Frequency Angular Momentum Biased Quasi-LTI Nonreciprocal Acoustic Filters," IEEE Trans. Ultrason. Ferroelectr. Freq. Control (UFFC), vol. 66, no. 11, July 2019, pp. 1814-1825.

[40] Y. Yu et al., "Radio Frequency Magnet-free Circulators Based on Spatiotemporal Modulation of Surface Acoustic Wave Filters," IEEE Trans. Microw. Theory Techn., IEEE Trans. Microw. Theory Techn., vol. 67, no. 2, pp. 4773-4782, 2019.

[41] Y. Yu et al., "Highly-Linear Magnet-Free Microelectromechanical Circulators," IEEE J. Microelectromech. Syst. (MEMS), vol. 28, no. 6, pp. 933-940, Dec. 2019.

[42] D. L. Sounas, N. A. Estep, A. Kord, and A. Alù, "Angular-Momentum Biased Circulators and Their Power Consumption," IEEE Antennas Wireless Propag. Lett., vol. 17, no. 11, Nov. 2018, pp. 1963-1967.

[43] A. Kord, D. L. Sounas, and A. Alù, "Differential Magnetless Circulator Using Modulated Bandstop Filters," in Proc. IEEE MTT-S Int. Microw. Symp. (IMS) Dig, Honolulu, HI, USA, Jun. 2017.

[44] A. Kord, D. L. Sounas, and A. Alù, "Low-Loss Broadband Magnetless Circulators for Full-Duplex Radios," in Proc. IEEE MTT-S Int. Mircow. Symp. (IMS), Philadelphia, Pennsylvania, USA, Jun. 2018.

[45] S. Qin, Q. Xu, and Y. E. Wang, "Nonreciprocal components with distributedly modulated capacitors," IEEE Trans. Microw. Theory Techn., vol. 62, no. 10, pp. 2260-2272, Oct. 2014.

[46] N. Reiskarimian and H. Krishnaswamy, "Magnetic-free non-reciprocity based on staggered commutation," Nat. Commun., vol. 7, Apr. 2016.

[47] N. Reiskarimian, J. Zhou, and H. Krishnaswamy, "A CMOS Passive LPTV Non-Magnetic Circulator and Its Application in a Full-Duplex Receiver," IEEE J. Solid-State Circuits, vol. 52, no. 5, pp. 1358-1372, May 2017.

[48] T. Dinc et al., "Synchronized conductivity modulation to realize broadband lossless magnetic-free non-reciprocity," Nat. Commun., vol. 8, no. 1, p. 795, Oct. 2017.

[49] A. Nagulu, A. Alù, and H. Krishnaswamy, "Fully-Integrated Non-Magnetic 180nm SOI Circulator with >1W P1dB, >+50dBm IIP3 and High Isolation Across 1.85 VSWR," in IEEE Radio Freq. Integ. Circuits Symp. (RFIC), Philadelphia, PA, USA, Jun. 2018.

[50] M. M. Biedka et al., "Ultra-Wide Band Non-reciprocity through Sequentially-Switched Delay Lines," Scientific Reports, vol. 7, Jan. 2017.

[51] M. M. Biedka et al., "Ultra-wide band on-chip circulators for full-duplex communications," in Proc. IEEE/MTT-S Int. Microwave Symp., 2018, pp. 987–990.

[52] C. Xu, E. Calayir, and G. Piazza, "Magnetic-free electrical circulator based on AlN MEMS filters and CMOS RF switches," in Proc. of the 31st IEEE Int. Conf. on Micro Electro-Mech. Sys. (MEMS 2018), Belfast, UK, Jan. 2018, pp. 755-758.

[53] Y. Yu et al., "Magnetic-Free Radio Frequency Circulator based on Spatiotemporal Commutation of MEMS Resonators," in Proc. of the 31st IEEE Int. Conf. on Micro Electro-Mech. Sys. (MEMS 2018), Belfast, UK, Jan. 2018, pp. 154-157.

[54] Y. Yu et al., "2.5 GHz Highly-Linear Magnetic-Free Microelectromechanical Resonant Circulator," in Proc. IEEE Int. Freq. Cont. Symp. (IFCS 2018), Olympic Valley, CA, May 2018.

[55] M. M. Torunbalci et al., "An FBAR Circulator," IEEE Microw. Wireless Compon. Lett., vol. 28, no. 5, pp. 395-397, May 2018.

[56] R. Fleury et al., "Sound isolation and giant linear nonreciprocity in a compact acoustic circulator," Science, vol. 343, pp. 516-519, Jan. 2014.

[57] Z. Yu and S. Fan, "Complete optical isolation created by indirect interband photonic transitions," Nat. Photon., vol. 3, pp. 91–94, Feb. 2009.

[58] H. Lira, Z. Yu, S. Fan, and M. Lipson, "Electrically driven nonreciprocity induced by interband photonic transition on a silicon chip," Phys. Rev. Lett., vol. 109, Jul. 2012, Art. no. 033901.

[59] K. Fang, Z. Yu, and S. Fan, "Photonic Aharonov–Bohm effect based on dynamic modulation," Phys. Rev. Lett., vol. 108, no. 15, Apr. 2012, Art. no. 153901.

[60] A. Kamal, J. Clarke, and M. H. Devoret, "Noiseless non-reciprocity in a parametric active device," Nat. Phys., vol. 7, no. 4, pp. 311–315, Apr. 2011.

[61] J. Kerckhoff et al., "On-Chip Superconducting Microwave Circulator from Synthetic Rotation," Phys. Rev. Appl., vol. 4, no. 3, Sept. 2015.

[62] K. M. Sliwa et al., "Reconfigurable Josephson Circulator/Directional Amplifier," Phys. Rev. X, vol. 5, no. 4, Nov. 2015.

[63] F. Lecocq et al., "Nonreciprocal Microwave Signal Processing with a Field-Programmable Josephson Amplifier," Phys. Rev. Appl., vol. 7, no. 2, Feb. 2017.

[64] A. M. Mahmoud, A. R. Davoyan, and N. Engheta, "All-passive nonreciprocal metastructure," Nat. Commun., vol. 6, no. 8359, July 2015.

[65] R. Duggan, D. L. Sounas, and A. Alù, "Optically Driven Effective Faraday Effect in Instantaneous Nonlinear Media," Optica, Vol. 6, No. 9, pp. 1152-1157, August 30, 2019.

[66] J. E. Moore, "The birth of topological insulators," Nature, vol. 464, no. 7286, pp. 194–198, Mar. 2010.

[67] M. Z. Hasan and C. L. Kane, "Colloquium: Topological insulators," Rev. Mod. Phys., vol. 82, no. 4, pp. 3045–3067, Nov. 2010.

[68] Z. Wang, Y. Chong, J. D. Joannopoulos, and M. Soljačić, "Observation of unidirectional backscattering-immune topological electromagnetic states," Nature, vol. 461, no. 7265, pp. 772–775, Oct. 2009.

[69] L. Lu, J. D. Joannopoulos, and M. Soljačić, "Topological photonics," Nat. Photonics, vol. 8, no. 11, pp. 821–829, Oct. 2014.

[70] R. O. Umucalılar and I. Carusotto, "Artificial gauge field for photons in coupled cavity arrays," Phys. Rev. A, vol. 84, no. 4, Oct. 2011.

[71] M. I. Shalaev, S. Desnavi, W. Walasik, and N. M. Litchinitser, "Reconfigurable topological photonic crystal," New J. Phys., vol. 20, no. 2, p. 023040, Feb. 2018.

[72] Q. Lin, X.-Q. Sun, M. Xiao, S.-C. Zhang, and S. Fan, "A three-dimensional photonic topological insulator using a two-dimensional ring resonator lattice with a synthetic frequency dimension," Sci. Adv., vol. 4, no. 10, p. eaat2774, Oct. 2018.

[73] K. Fang, Z. Yu, and S. Fan, "Realizing effective magnetic field for photons by controlling the phase of dynamic modulation," Nat. Photonics, vol. 6, no. 11, pp. 782–787, Oct. 2012.

[74] M. Hafezi, S. Mittal, J. Fan, A. Migdall, and J. M. Taylor, "Imaging topological edge states in silicon photonics," Nat. Photonics, vol. 7, no. 12, pp. 1001–1005, Dec. 2013.

[75] M. Pasek and Y. D. Chong, "Network models of photonic Floquet topological insulators," Phys. Rev. B, vol. 89, no. 7, Feb. 2014.

[76] W.-J. Chen et al., "Experimental realization of photonic topological insulator in a uniaxial metacrystal waveguide," Nat. Commun., vol. 5, no. 1, Dec. 2014.

[77] T. Ma, A. B. Khanikaev, S. H. Mousavi, and G. Shvets, "Guiding Electromagnetic Waves around Sharp Corners: Topologically Protected Photonic Transport in Metawaveguides," Phys. Rev. Lett., vol. 114, no. 12, Mar. 2015.

[78] X. Cheng, C. Jouvaud, X. Ni, S. H. Mousavi, A. Z. Genack, and A. B. Khanikaev, "Robust reconfigurable electromagnetic pathways within a photonic topological insulator," Nat. Mater., vol. 15, no. February, pp. 1–8, 2016.

[79] X. Ni, M. Weiner, A. Alù, and A. B. Khanikaev, "Observation of higher-order topological acoustic states protected by generalized chiral symmetry," Nat. Mater., vol. 18, no. 2, pp. 113–120, Feb. 2019.

[80] A. B. Khanikaev, R. Fleury, S. H. Mousavi, and A. Alù, "Topologically robust sound propagation in an angular-momentum-biased graphene-like resonator lattice," Nat. Commun., vol. 6, no. 1, Dec. 2015.

[81] R. Fleury, A. B. Khanikaev, and A. Alù, "Floquet topological insulators for sound," Nat. Commun., vol. 7, p. 11744, Jun. 2016.

[82] A. Darabi, X. Ni, M. Leamy, and A. Alù, "Reconfigurable Floquet Elastodynamic Topological Insulators Based on Synthetic Angular Momentum Bias," Science Advances, 6, eaba8656, July 2020.

[83] C. H. Lee et al., "Topolectrical Circuits," Commun. Phys., vol. 1, no. 1, Dec. 2018.

[84] S. Imhof et al., "Topolectrical-circuit realization of topological corner modes," Nat. Phys., vol. 14, no. 9, pp. 925–929, Sep. 2018.

[85] M. Tymchenko and A. Alù, "Circuit-based magnetless floquet topological insulator," in Proc. 10th International Congress on Advanced Electromagnetic Materials in Microwaves and Optics (METAMATERIALS), 2016.

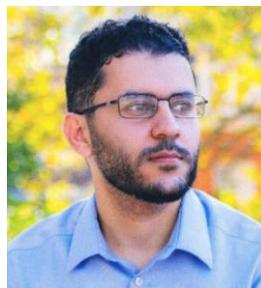

**Ahmed Kord** (GS'15) received the B.S. and M.S. degrees from Cairo University, Egypt, in 2011 and 2014, respectively, and the Ph.D. degree from the University of Texas at Austin, TX, USA, in 2019, all in electrical engineering. In 2015 and 2018, he was with Qualcomm Inc. and Intel Labs, respectively, as a summer intern. From Dec. 2016 to Aug. 2017, he was with Eureka Aerospace Inc. as a part-time consultant. From May 2019 to June 2020, he was with Columbia University as a postdoctoral research scientist. He is currently with Qualcomm Inc. as a senior RFIC design engineer. In Dec. 2019, he was named to the Forbes 30 Under 30 in Science.

Dr. Kord is a co-author of more than 30 scientific papers and a co-inventor of two patents, one of which is licensed to the US startup company Silicon Audio/RF Circulator. He is also a recipient of several prestigious awards including the Qualcomm Innovation Fellowship, the IEEE MTT-S graduate fellowship, the IEEE AP-S doctoral research award, the graduate student excellence award, the graduate dean's fellowship, the Douglas Wilson fellowship, the undergraduate student excellence award, and several travel fellowships and first place awards in different student paper, poster and design competitions.

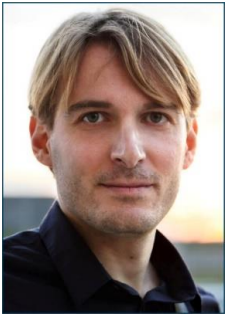

**Andrea Alù** (Fellow 2014, IEEE) is the Founding Director of the Photonics Initiative at the CUNY Advanced Science Research Center. He is also the Einstein Professor of Physics at the CUNY Graduate Center, Professor of Electrical Engineering at the City College of New York, and Adjunct Professor and Senior Research Scientist at the University of Texas at Austin. He received the *Laurea*, MS and PhD degrees from the University of Roma Tre, Rome, Italy, respectively in 2001, 2003 and 2007. After spending one year as a postdoctoral research fellow at UPenn, in 2009 he joined the faculty of the University of Texas at Austin, where he was the Temple Foundation Endowed Professor until 2018. His current research interests span over a broad range of areas, including metamaterials and plasmonics, electromangetics, optics and nanophotonics, acoustics, scattering, nanocircuits and nanostructures, miniaturized antennas and nanoantennas, RF antennas and circuits. He is a Fellow of the National Academy of Inventors, AAAS, IEEE, OSA, APS and SPIE, a Highly Cited Researcher since 2017, a five times Blavatnik awardee and a Simons Investigator in Physics. He has received several awards for his scientific work, including the IEEE Kiyo Tomiyasu award (2020), the NSF Alan T. Waterman award (2015), the Vannevar Bush Faculty Fellowship from DoD (2019), the ICO Prize in Optics (2016), the OSA Adolph Lomb Medal (2013), and the URSI Issac Koga Gold Medal (2011).